\documentclass{amsart}
\usepackage{amsfonts}
\usepackage{blkarray}
\usepackage{float}
\usepackage{url}
\usepackage{graphicx}
\usepackage{multirow}
\usepackage{algorithm}
\usepackage[noend]{algpseudocode}

\theoremstyle{definition}

\theoremstyle{remark}

\numberwithin{equation}{section}

\makeatletter
\def\BState{\State\hskip-\ALG@thistlm}
\makeatother

\begin{document}
\title[Blakley’s Secret Sharing Scheme]{Overview of Blakley’s Secret Sharing Scheme}
\author{Alireza Shamsoshoara}
\address{School of Informatics, Computing, and Cyber Systems, Department of Mathematics \& Statistics, Northern Arizona University, Flagstaff, AZ 86001
 \& }
\email{alireza\_shamsoshoara@nau.edu}

\begin{abstract}
In this report, I explained the problem of Secret Sharing Scheme. Then based on the definition of the problem, two old methods: Blakley's Secret Sharing Scheme and Shamir's Secret Sharing are introduced. However, we explained the details of the first one since it's the topic of this work. Blakley's method has an application in distributing a key between different parties and reconstructing the key based on each share. However, this method is not efficient enough because of too large space states. Also, we tried to simulate a scenario for spreading a key between some users and tried to reconstruct the first key using Matlab in a graphical user interface.
\end{abstract}

\maketitle

\section{Introduction}

I started the problem of secret sharing scheme as a classical problem based on \cite{liu1968introduction}. Assume that there are eleven scientists and they are working on a very confidential project. Their purpose is to hide the documents and results of the project in a safe box in a way that the safe cannot be opened unless six or more scientists are available in the lab. So the question is how the head of the group can spread a common key between scientists that no one can make the original key with his/her individual key, but a combination of scientists can retrieve the key to open the safe box.

In another scenario assume that Coca Cola executives want to have access to ``Secret Formula" in case of the emergency. So they can access the ``Secret Formula" if they have these three possible combinations: i) six Directors or ii) three vice presidents or iii) one president. So the question is how one reliable person can spread and announce the ``Secret formula" between these parties that those combinations can reconstruct the formula again in the case of the emergency.

In another case, consider the Rivest Shamir Adleman (RSA) \cite{rivest1978method}. RSA is mainly used in crypto-systems for secure transmission between two entities in the network. In this mechanism, everyone has a private key and public key. Private key is secret and it is only used for the decryption of the encrypted message at each user; however, public key is being used for encrypting the message. The decryption key is also called private key. This method of encryption and decryption is called asymmetric, since, the private and public (decryption and encryption) keys are different. This method is based on choosing two different large prime numbers which makes the public key. Anyone can utilize the public key to encrypt its own message. The values of those two large prime numbers should be kept secret from everyone.
But with currently published methods, and if the public key is large enough, only someone with knowledge of the prime numbers can decode the message feasibly \cite{rivest1978method}. Breaking RSA encryption is known as the RSA problem. Whether it is as difficult as the factoring problem remains an open question. 
So another application of the secret sharing scheme is if one user or entity wants to trust another user to share the private key with him/her, how can he/she share that key? Is it enough to share with just one or it is better to share the private key with multiple persons in such a way that no one can predict the key from its share.

\begin{figure}[t]
	\centering
	\includegraphics[width=.8\columnwidth]{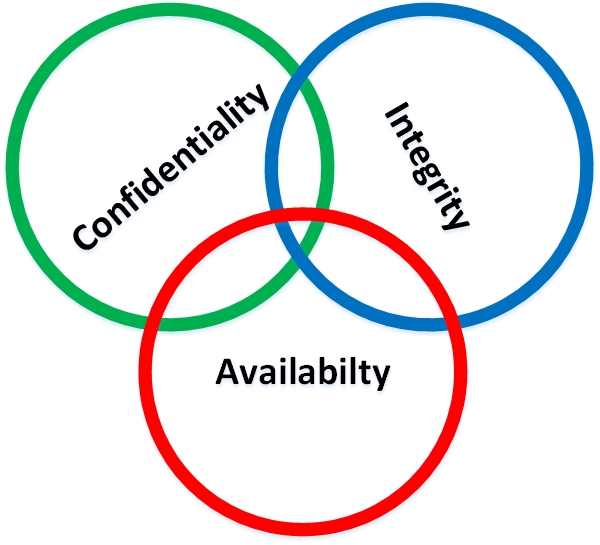}
	\caption{Three main criteria in secret sharing scheme}
	%\vspace{-0.15in}
% 		\vspace{-10pt}
    \label{fig:Fig1}
\end{figure}

In all these three scenarios, there is one common objective: Is it possible to store data among multiple
semi-trusted people/nodes in a manner that doesn't violate any of the three cornerstones of information security? Figure \ref{fig:Fig1} demonstrates these three important criterion. The answer of this question is yes. Schemes such as Blakley\cite{blakley1979safeguarding} and Shamir\cite{shamir1979share} are popular for replying to these challenges and scenarios. 

% *****************************************
% *****************************************

\subsection{Definition}
In this subsection, we will define the secret sharing scheme based on Blakley's and Shamir's secret sharing schemes. In addition to Blakley and Shamir's secret sharing schemes, there is another method based on Chinese Remainder Theorem (CRT) which has been introduced by Asmuth and Bloom in 1983 \cite{asmuth1983modular}. Any method for distributing a secret among a group of individuals \textbf{(shareholders)} each of which is allocated some information \textbf{(share)} related to the secret. These two items should be common between in all these three methods:
\begin{itemize}
    \item The secret can only be reconstructed when the
        shares are combined together.
    \item Individual shares are of no use on their own
\end{itemize}

Moreover, these schemes are also called $(t, n)$ threshold schemes. since the secret is distributed among $n$ participants and only $t$ or more participants can recover the secret. The goal of these schemes is to divide a secret $S$ into $n$ shares $S_0, \dotsm, S_{n-1}$ such that:
\begin{enumerate}
    \item knowledge of $t$ or more shares makes S easily
        computable
        \item knowledge of $t - 1$ or less shares leaves $S$ completely undetermined
\end{enumerate}

where:
\begin{itemize}
    \item $n$ and $t$ are positive integers.
    \item $n$ is the number of secrets generated.
    \item $t$ is the amount of overall secrets needed to recover the original secret.
    \item $t$ is less than or equal to $n$.
\end{itemize}

Now if we want to investigate the secrecy, integrity, and availability we can summarize the results in these bullets: 
\begin{itemize}
    \item \textbf{Secrecy:} If an eavesdropper wants to learn the secret, it needs to corrupt at least $t$ shareholders and collect their shares.
    \item \textbf{Integrity:} If an eavesdropper wants to destroy or alter the secret, it needs to corrupt at least $n - t + 1$.
    \item \textbf{Availability:} If the threshold value ($t$) is known and given, then the secret availability will be increased as $n$ increases.
    \\
    If the value of $n$ (number of shareholders) increases, then the secrecy and integrity will be enhanced if $t$ increases.
\end{itemize}

\section{Blakley's Secret Sharing Scheme}
In this scheme, the dealer who knows the secret, spreads the key or secret among $n$ members. Each member would call his or her piece as share. The key will be revealed if $t$ or more of them share the information together. Blakley utilized a geometric approach to solve this problem. He assumed that the secret is a point in a $t$-dimensional space. Based on this $t$-dimensional space, if hyperplanes intersect at one point, it will reconstruct the secret key. Coefficients of $n$ different hyperplanes constitute the corresponding $n$ shares. This scheme is a linear threshold scheme like Shamir's scheme. 

\begin{figure}[!hbt]
	\centering
	\includegraphics[width=.5\columnwidth]{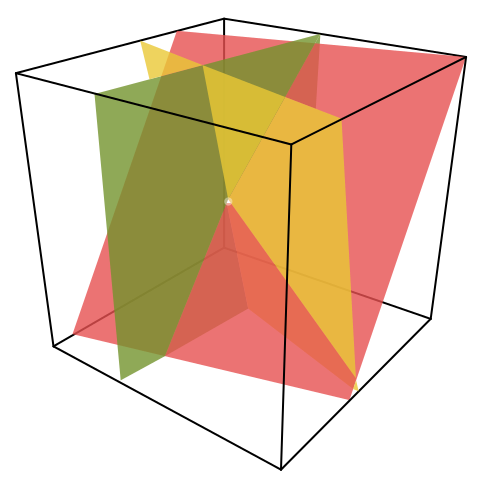}
	\caption{Three different hyperplanes (shareholders) \cite{FileSecr71online}}
	%\vspace{-0.15in}
% 		\vspace{-10pt}
    \label{fig:hyperplane}
\end{figure}

In Blakley's approach, the secret and the shares can be summarized as a linear system $Cx = y$, where matrix $C$ and the vector $y$ correspond to hyperplane equations.
Blakley’s method \cite{blakley1979safeguarding} utilized principles of geometry to share the secret. According to this method, the secret key is a point in a $t$-dimensional space, which is the intersect point of all the hyperplanes. Affine hyperplanes in this space represent n shares. Blakley secret sharing scheme can be represented as a linear system $Cx$ mod $p = y$. The general full rank matrix C is the critical data in this approach. In the following, we will explain the two main concepts (distributing and reassembling) in this approach. At first, consider Fig \ref{fig:hyperplane} which demonstrates three different hyperplanes in 3-Dimensional space. Each plane represents one shareholder based on the definition. 

% *****************************************
% *****************************************

% *****************************************
\begin{figure}[!hbt]
	\centering
	\includegraphics[width=\columnwidth]{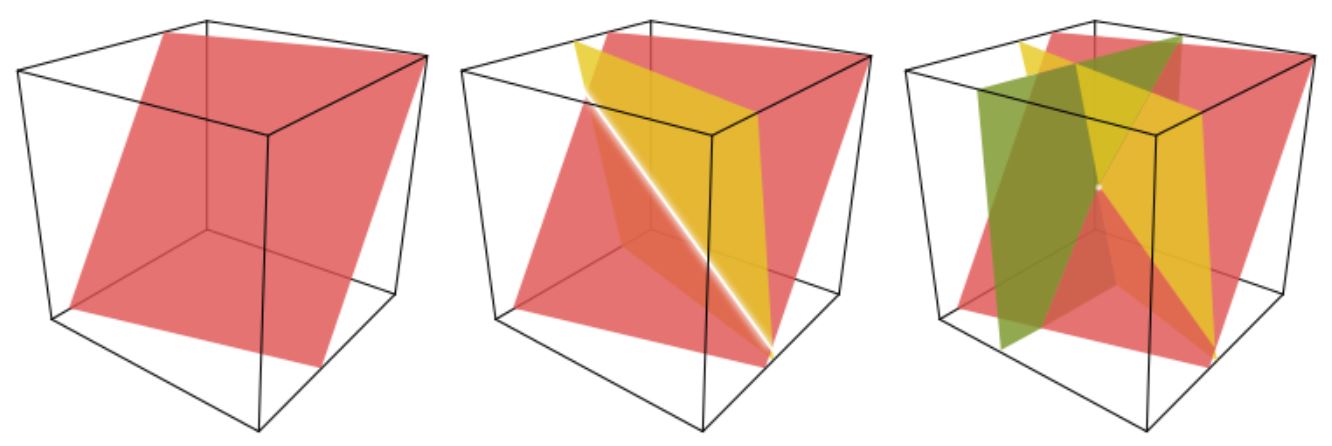}
	\caption{Intersection of different hyperplanes \cite{FileSecr71online}}
	%\vspace{-0.15in}
% 		\vspace{-10pt}
    \label{fig:hyperplanes}
\end{figure}
% *****************************************

As it is demonstrated in Fig\ref{fig:hyperplanes}, intersection of three and two hyperplanes define a line and a point in a 3-Dimentional space respectively. The main idea is two nonparallel lines in the same plane intersect exactly at one point. Three nonparallel planes in space intersect exactly at a specific point. Any $t$ nonparallel $(t - 1)$ dimensional hyperplanes intersect at a specific point. The secret may be encoded as any single coordinate of the of the intersection. If the key is encoded using all the coordinates, then the insider will gain information about the key because he knows the point must lie on his plane. If the insider can access more information about others too, then the system in not
safe anymore.

In the following, I will itemize the process of distributing the secret shares between shareholders. This task can be done by the dealer. 
\begin{enumerate}
    \item Pick a prime number ($p$)
    \item Create a point using your secret information     $x_0$
    \item Pick random values for the $Y_0$ and $Z_0$ axes in $\textrm{mod}$ $p$
    \item Use previous values to make your intersection point $Q(x_0, y_0, z_0)$
    \item Choose values for $a$ and $b$ in $\textrm{mod}$\ $p$ and find the value for your $c$ in your hyperplane such that:
    \begin{align}\label{eq:1}
        c \equiv z_0 - ax_0 - by_0 \ (\textrm{mod}\ p)
    \end{align}
    \item Based on the values of $a$, $b$, and $c$, define your hyperplane:
    \begin{align}\label{eq:2}
        z \equiv ax + by + c \ (\textrm{mod}\ p)
    \end{align}
\end{enumerate}

Now everyone has a specific hyperplane. For instance, making hyperplanes for five share holders will result to:
\begin{align}\label{eq:3}
    &a_1x + b_1y - z \equiv -c_1 \ (\textrm{mod} p) \\
    \nonumber&a_2x + b_2y - z \equiv -c_2 \ (\textrm{mod} p) \\   
    \nonumber&a_3x + b_3y - z \equiv -c_3 \ (\textrm{mod} p) \\
    \nonumber&a_4x + b_4y - z \equiv -c_4 \ (\textrm{mod} p) \\
    \nonumber&a_5x + b_5y - z \equiv -c_5 \ (\textrm{mod} p)
\end{align}

which can be summarized as: 
\begin{align}\label{eq:4}
    &a_ix + b_iy - z \equiv -c_i \ (\textrm{mod} p) \qquad 1 \leq i \leq 5
\end{align}

Hence, choosing only three shareholder from five people would yield to find the secret key which is $x_0$ in this specific example. Hence, finding solution for equation \ref{eq:5} will find the value for the secret key. As long as determinant of matrix is nonzero in $\textrm{mod} \ p$, matrix can be inverted and the secret can be found.

\begin{align}\label{eq:5}
    \left(\begin{array}{ccc} a_1 & b_1 & -1\\
                             a_2 & b_2 & -1\\
                             a_3 & b_3 & -1\end{array} \right)
    \left(\begin{array}{c}   x_0\\
                             y_0\\
                             z_0\end{array} \right)
    \equiv 
    \left(\begin{array}{c}   -c_1\\
                             -c_2\\
                             -c_3\end{array} \right)
                             \
                             \textrm{mod} \ (p)
\end{align}

Subsection \ref{subsec:example} gives one example of this matrix representation and these two main concepts.

Two main drawbacks of this schemes are:
\begin{itemize}
    \item Blakley’s scheme lacks actual implementations. In \cite{blakley1994linear}, Blakley et al. only presented a guideline to outline a matrix of linear systems for perfect secrecy, and no actual matrix was indicated.
    \item This approach is not space efficient.
    \item Each shareholder knows that the point (which is the secret) lies in his share (hyperplane). As a consequence, the method is not perfect. 
\end{itemize}
Maybe, because of these two reasons, Shamir's scheme is more popular compared to the Blakley's approach.

In the following, I summarize some details of recent works which are related to this scheme. Recently, researchers started to use Blakley’s geometry-based secret sharing approach in the area of secret image sharing \cite{chen2008geometry, tso2008sharing}. Ulutas et al. \cite{ulutas2009improvements} introduced an imprved scheme for secret image sharing, which adopts Blakley’s secret sharing method with another method to share the secret and create meaningful shares. Moreover, Bozkurt et al.\cite{bozkurt2008threshold} explained the first threshold RSA signature approach using the Blakley's secret scheme.

\subsection{Example of Blakley's Secret Sharing Scheme}
\label{subsec:example}

As an example for equations in the previous section consider this case. Assume that there are five different hyperplanes. Also, assume that $p$ is equal to 73 which is prime. 

\begin{align}\label{eq:6}
    &z \equiv 4x \ + 19y + 68 \ (\textrm{mod} 73) \\
    \nonumber&z \equiv 52x + 27y + 10 \ (\textrm{mod} 73) \\   
    \nonumber&z \equiv 36x + 65y + 18 \ (\textrm{mod} 73) \\
    \nonumber&z \equiv 57x + 12y + 16 \ (\textrm{mod} 73) \\
    \nonumber&z \equiv 34x + 19y + 49 \ (\textrm{mod} 73)
\end{align}

solving equation \ref{eq:7} would find the secret key as $42$.

\begin{align}\label{eq:7}
    \left(\begin{array}{ccc} 4 & 19 & -1\\
                             52 & 27 & -1\\
                             36 & 65 & -1\end{array} \right)
    \left(\begin{array}{c}   x_0\\
                             y_0\\
                             z_0\end{array} \right)
    \equiv 
    \left(\begin{array}{c}   -68\\
                             -10\\
                             -18\end{array} \right)
                             \
                             \textrm{mod} \ (73)
\end{align}

\vspace{3mm}
The solution is $(x_0, y_0, z_0)$ = $(42, 29, 57)$ which means secret key = $x_0$ = $42$

\section{Using Graphical User Interface (GUI) in Matlab to find hyperplanes and solutions}

In this section, I designed a program for users to assign the values for the initial prime number, choose numbers of shareholder, number of selected shareholders, and the private key. A user can define these values in the application. Then based on the algorithm in the \ref{algorithm}, the program calculates the coefficient matrix and the vector with constant values. For solving the linear equation, $gflineq$ command is used in $\textrm{mod} \ p$.

\begin{algorithm}
\caption{Blakley's secret sharing scheme Pseudo code}\label{algorithm}
\begin{algorithmic}[1]
\Procedure{Blakley}{}
\State $\textit{prime number} \gets \text{user Input}$
\State $\textit{Secret Key} \gets \text{user Input}$
\State $\textit{Number of Shareholders} \gets \text{user Input}$
\State $\textit{Number of selected Shareholders} \gets \text{user Input}$
\BState \emph{loop}:
\State $\textit{Random point} \gets \text{Random integers between 0 and P - 1}$
\State $\textit{Random coefficients} \gets \text{Random integers between 0 and P - 1}$
\State $\textit{C} \gets \text{mod(z - ax - by, P)}$
\State $\textit{Matrix} \gets \text{[a, b, mod(-1, P)]}$
\If {$\textit{det}(Matrix) = 0$}
\State \textbf{goto} \emph{loop}.
\EndIf
\State $\textit{Selected Random} \gets \text{random permuation(share holder, selected)}$

\State $\textit{Solution} \gets \text{solving linear equation for Galois Field for selected rows)}$

\EndProcedure
\end{algorithmic}
\end{algorithm}

\begin{figure}[!hbt]
	\centering
	\includegraphics[width=\columnwidth]{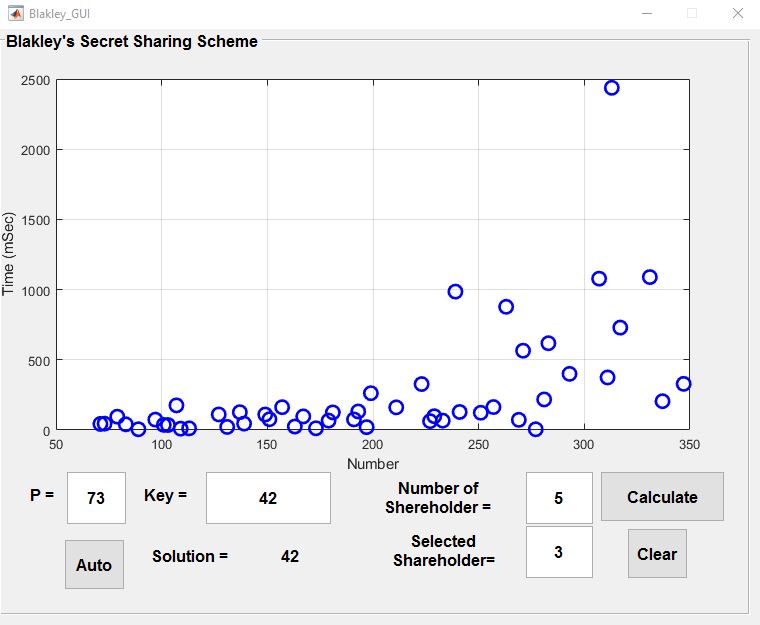}
	\caption{Time simulation for the Blakley's secret sharing scheme}
	%\vspace{-0.15in}
% 		\vspace{-10pt}
    \label{fig:simulation}
\end{figure}

After running the program, every output demonstrates one solution of a point which includes the secret key too. Each time we try to stamped the time to see what is the effect of each prime number on the computation time. It seems that the growing rate for the time is exponential when the initial prime number is getting big. Fig \ref{fig:simulation} demonstrates this behavior for first 100 prime numbers with 5 share holders and 3 selected shares.  

% *****************************************
% *****************************************

\section{Conclusion}
In this article, I reviewed the secret sharing schemes specifically Blakley approach. Then I tried to simplify the problem by using an example and the graphical user interface in the Matlab. we proved that choosing large prime numbers will take more time for the algorithm to solve the linear equation.

% \begin{thebibliography}{99}
% \bibitem{betsumiya} K. Betsumiya, S. Ling and F.R. Nemenzo,
% \textquotedblleft Type II codes over $\mathbb{F}_{2^{m}}+u\mathbb{F}_{2^{m}}$%
% ", Discrete Mathematics, Vol. 275 pp. 43-65, 2004.

% \bibitem{ling} S. Ling and P. Sole, \textquotedblleft Type II codes over $%
% \mathbb{F}_{4}+u\mathbb{F}_{4}$", Europ. J. Combinatorics, Vol. 22
% pp. 983--997, 2001.

% \bibitem{macwilliams} F.J. Macwilliams, N.J.A. Sloane, The theory of error
% correcting codes, Amsterdam: North-Holland, 1977.

% \bibitem{Rains} E.M. Rains, \textquotedblleft Shadow Bounds for Self Dual
% Codes", IEEE Trans. Inform. Theory, Vol. 44, pp.134--139, 1998.

% \end{thebibliography}.

\bibliographystyle{plain}
\bibliography{sample.bib}

\end{document}